\documentclass[pre,unsortedaddress,twocolumn,showpacs]{revtex4}

\usepackage{graphicx}
\usepackage{amsmath}

\begin{document}
\newcommand{\nn}{\nonumber}

\bibliographystyle{prsty}                                          
\title{Peptide size dependent active  transport in the proteasome}

\author{Alexei Zaikin and Thorsten P\"oschel}
\affiliation{Institut f\"ur Biochemie, Humboldt Universit\"at zu Berlin, Charit\'e,\\ Monbijoustra{\ss}e 2, 10117 Berlin, Germany}

\date{\today}

\begin{abstract}
We investigate the transport of proteins inside the proteasome and propose an active transport mechanism based on a spatially asymmetric interaction potential of peptide chains. The transport is driven by fluctuations which are always present in such systems. We compute the peptide-size dependent transport rate which is essential for the functioning of the proteasome. In agreement with recent experiments, varying temperature changes the transport mechanism qualitatively.
\end{abstract}


\maketitle

Eukaryotic mammalian proteasomes are fascinating molecular machines, intended to digest tagged with ubiquitin proteins \cite{2001_Goldberg_sa}. The proteins are cleaved into peptides which can be further used for the synthesis of new  proteins. The proteasome's function has been directly linked to a wide range of diseases, such as cancer, neurological diseases, and diseases of the immune defense system. Starting from the discovery of their activity in the late 1970's, presently  proteasomes are in the focus of current molecular biology, see \cite{HiltWolf:2001CiechanoverMasucci:2003} and  refs. therein. By now there are several models to explain the proteasome's activity \cite{2003_Schliwa_nature}, cleavage mechanisms \cite{2000_Altuvia_jmb,1999_Holzhuetter_jmb,2002_Kesmir} and to predict the cleavage results \cite{1999_Kisselev_jbc}, however, basic principles of the proteasome operation mechanisms are still poorly understood, mainly due to the lack of experimental results.

In this Letter, we focus on understanding of protein translocation inside the proteasome, leaving  the mechanisms of cleavage, targeting, etc. beyond the scope. The main question is: given proteasomes are highly complex pipe-like structures of tenthousands atoms of almost perfect left-right symmetry with respect to the axis of the pipe \cite{1998_Walz}, thus, operating bi-directional. To be cleaved, a protein has to enter the proteasome at one side, pass the active sites (where the cleavage occurs), which makes about 1/3 of the total length of the pipe. Then the cleaved peptides have to pass all the way through the pipe to finally reappear at the other side of the proteasome. Although there is a number of examples where protein transport in cells occurs due to diffusion, i.e. Brownian motion \cite{2001_Reits_nature_cell}, diffusive transport may be excluded as the main mechanism for translocation in the proteasome because of the enormous cargo \cite{2000_Elston_siam}.  Also other proposed transport mechanisms, such as the power stroke model of protein translocation do not suffice to explain translocation \cite{2002_Elston_bio_journal}. Therefore, the question arises how the protein's motion inside the proteasome is driven?

Since proteasomes are large multi-subunit structures consisting of proteins, the mechanism of the protein transport is directly related to protein-protein interaction. In \cite{2001_Brokaw_bio_journal} it has been noticed that if attachment and detachment rates are specified asymmetrically, the protein-protein binding interaction acts as a ratchet. Following this argumentation, active protein transport, based on the mechanism of a molecular ratchet, has been studied for transport through membranes \cite{1990_Vale1994_Astumian_prl2002_Reimann_pr} and has been also discussed as a mechanism for cytosolic destruction by the proteasome \cite{1999_Matlack_cell}. Moreover, maximum
likelihood tests have shown that other models, e.g. the power stroke model of protein translocation,
do not lead to better agreement with  the experiment than the Brownian ratchet model \cite{2002_Elston_bio_journal}. Noteworthy, in these ratchet effects transport is possible only in a certain temperature interval, and stochasticity, intrinsically present due to fluctuations in any biochemical reaction \cite{1981_van_Kampen_book,1999_Barkai_nature}, provides the driving mechanism. 

Here we propose a model for active protein translocation in the proteasome to explain the peptide size dependence of the transport velocity as well as its temperature dependence which possibly explains the mechanism of temperature reaction or  heat shock response, regulating the proteasome activity in the case of  some diseases \cite{2000_Kuckelkorn,2000_Wyttenbach_pnas,2001_Pritts,2001_Stepanenko_mol_biology}. The results describe a system size ratchet effect that is  related to similar effects which have been described recently for other noise-induced phenomena \cite{2002_Pikovsky_prl}. It has been shown {\em in vitro} (e.g. \cite{HiltWolf:2001CiechanoverMasucci:2003}) that already the 20S proteasome reveals full functionality, thus, for studying the transport we disregard the enzymatic action of the proteasome caps.

\begin{figure}[t!]
  \centerline{\hspace*{0.3cm}\includegraphics[width=6.5cm,clip]{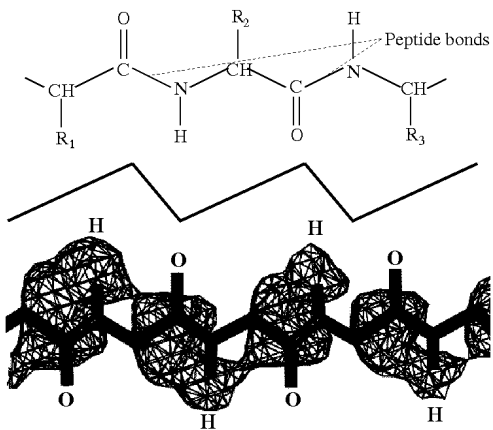}}
  \caption{The regular part of the protein-proteasome interaction potential is modeled by a  spatially asymmetric periodic potential, based on the $C-N$ asymmetry of the peptide backbone. The $R_i$ mark the amino-acid-specific residues. The lower part shows the equipotential surfaces of a simple peptide which is clearly asymmetric (see also \protect\cite{asymmetry}). The $H$ and $O$ atoms are marked for orientation.}
  \label{potential}
\end{figure}

{\em Model.} 
Following, e.g., \cite{2001_Brokaw_bio_journal} we assume that the protein-proteasome interaction is characterized by a spatially periodic asymmetric potential, motivated by asymmetric C-N binding between amino acids that enter the proteasome {\em always} in $N\to C$ direction, i.e., with $N$ head on \cite{1999_Holzhuetter_jmb} (see Fig. \ref{potential}). The folding structure of the protein is not relevant here since due to the action of the regulatory complexes  the protein enters the proteasome unfold \cite{1997_Kopito_cell}. Moreover, even fold proteins tends to form periodic structures to maximize their amphiphilicity \cite{1984_Eisenberg_pnas}. 

As an abstraction, here we assume that the protein-proteasome interaction potential $U(x)$ is periodic with the period $L$. In reality there is a basic asymmetry, namely the $C-N$ asymmetry of the protein (or peptide) backbone, that is superposed by a non-periodic (in our sense irregular) part that is attributed to the amino-acid-specific residues \cite{asymmetry}. 
The basic structure of a peptide and its model as an asymmetric ratchet potential is sketched in Fig. \ref{potential}. The commensurability of the protein and the proteasome is supported by the fact that both biological macro-molecules consist of the same basic structures, namely chains of amino-acid. That commensurability has been claimed before, and was the basis of automaton-like models of proteasome digestion processes \cite{1999_Holzhuetter_jmb}.
The detailed form of the asymmetric periodic interaction potential is of less importance for this qualitative study. Here, we assume a saw-tooth potential as drawn in Fig. \ref{potential}. The angles are smoothed, i.e., $dU/dx$ can be computed in each point (for details see \cite{2001_Landa_Zaikin_csf}).

\begin{figure}[t!]
\centerline{\includegraphics[width=0.475\textwidth]{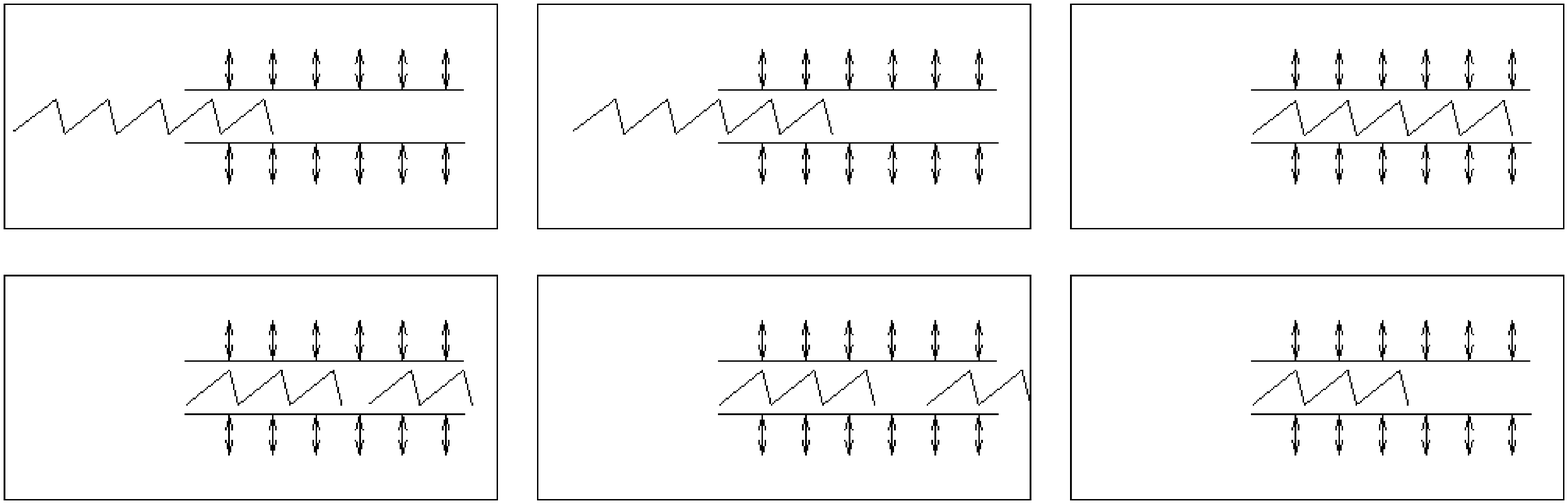}}

  \caption{From left to  right; Top:  protein moves into the proteasome while increasing the number $N$ of interaction centers. Bottom: after cleavage $N$ abruptly decreases for the resulting peptides. When leaving the proteasome, $N$ decreases as well.
}
\label{work}
\end{figure}

The proteasome acts upon the protein by a certain number of equidistant interaction centers. The dynamics of the protein inside the proteasome is, hence,  governed by $N$ interactions centers, where $N$ is the number of protein elements (amino acids or multiplicatives of it). 
There appear the following forces: potential force (protein-proteasome interaction)  $-N \partial U(x)/\partial x$, 
fluctuations with collective $N F(t)$ and individual components $f_1(t)+...+f_N(t)$, and protein friction forces $N \beta \dot x$ \cite{2001_Brokaw_bio_journal},
 where $x$ is the coordinate of the
protein with respect to the proteasome and $\beta$ is the coefficient of friction. 
Due to the small mass of all protein particles, moving in the liquid cytosol, 
the motion occurs in the overdamped realm \cite{2003_Bier_prl}, hence we  neglect inertia forces.
Note that transport is possible only in the case of nonequilibrium
fluctuations.
In the simplified case, when  fluctuations can be represented by a sum of a collective periodic force 
and individual for every protein residue  thermal noise, 
the model is analytically tractable, predicting the velocity dependence on the peptide size.
Numerically, we investigate also different kinds of fluctuations 
in order to be closer to  reality.
Normalizing all forces by friction and taking $\beta=1$, the translocation of a protein in the proteasome is governed then by 
\begin{eqnarray}
\frac{\partial x}{\partial t} = - \frac{\partial U(x)}{\partial x}+F(t)+\frac{1}{N}(f_1(t)+...+f_N(t)).
\label{eq1}
\end{eqnarray}
When the protein chain enters the proteasome, the number of interaction centers $N$ is increased (Fig.~\ref{work}). 
After cleavage, $N$ is abruptly decreased for the cleavage products.
We believe that this crucially changes the transport velocity, providing the cleavage product leaving the proteasome. Eq.\eqref{eq1} models the motion for both the initial protein and the peptides after cleavage. We consider three cases of fluctuations, regulating their motion:

{\it Case 1:} We assume collective oscillations of the peptide elements \cite{1997_Julicher_prl,2000_Tama_proteins}, e.g., $F(t)=A\cos(\omega t)$, where $A$ and $\omega$ stand for the amplitude and frequency of these oscillations. Additionally, each interaction center undergoes local thermal fluctuations, represented by  mutually uncorrelated white noise of intensity $\sigma^2$: $f_i(t)=\xi_i(t)$, where 
$\langle \xi_i(t) \xi_j(t^{\prime})\rangle = \sigma^2 \delta (t-t^{\prime})\delta_{ij}$. In this case the stochastic term in Eq. \eqref{eq1} is white noise of intensity $\sigma^2/N$. 
The Fokker-Planck equation for the peptide coordinate probability distribution $w(x,t)$
associated with Eq. \eqref{eq1} is
\begin{equation}
\frac{\partial w}{\partial t} = -\,\frac{\partial}{\partial x}\left[\left(F (t)-\frac{\partial U}{\partial x}
\right)w(x,t )\right]+ \frac{\sigma^2}{2 N}\,\frac{\partial^2w(x,t)}{\partial x^2},
\nonumber
\end{equation}
which may be solved in quasi-stationary adiabatic approximation  $\partial w/\partial t=0$ \cite{1990_Jung_pra}. We  obtain 
\begin{equation}
\label{2.6}
\frac{\sigma^2}{2 N}\,\frac{\partial w(x,F )}{\partial x}- 
\left(F -\frac{\partial U}{\partial x}\right)w(x,F )=-G(F ),
\end{equation}
where $G(F )$ is the probability flux. For any periodic potential $U(x)$
the quasi-stationary solution of Eq. \eqref{2.6} is
\begin{equation}
\nonumber
\begin{split}
w(x,t)=&\left[C(F )-\frac{2G(F )}{\sigma^2/N}\int_0^x 
\exp\left(\frac{U(x^{\prime})-F x^{\prime}}{\sigma^2/2N}\right)\,dx^{\prime}\right]\\
&\exp\left(-\,\frac{U(x)-F x}{\sigma^2/2N}\right),
\end{split}
\end{equation}
where $C(F (t))$ and $G(F (t))$ are unknown functions of $t$.
Using  the periodicity condition $w(0,t)=w(L,t)$  and the normalization of $w(x,t)$ we get $G(F)$.
If the amplitude $A$ meets the condition 
$LA\ll \sigma^2/N,$
one can expand $G(F)$ and obtain 
\begin{equation}
\label{3.16}
 G(F ) \approx G_{01}F +G_{02}F ^2
\end{equation}
with the  expansion coefficients $G_{01}=L/(I_{10}I_{20})$,
\begin{equation}
  \begin{split}
G_{02}&=G_{01}\left(\frac{I_{11}}{I_{10}}-\frac{I_{21}}{I_{20}}-\frac{N L}{\sigma^2}\left(1-\frac{2I_{30}}{ I_{10}I_{20}}
\right)\right)\\
I_{10}&=\int_0^{L}e^{U^\prime(x)}\,dx,\,~~ I_{20}=\int_0^{L}e^{\left(-U^\prime(x)\right)}\,dx\\
I_{11}&=\frac{2N}{\sigma^2}\int_0^{L}xe^{U^\prime(x)}\,dx,\,\, I_{21}=\frac{2N}{\sigma^2}\int_0^{L}xe^{\left(-U^\prime(x)\right)}\,dx\\
I_{30}&=  \int_0^{L}\int_0^{x}
e^{\left( U^\prime(x^\prime)-U^\prime(x)\right)}\,dx^{\prime}\,dx,\,U^\prime(x)=\frac{2NU(x)}{\sigma^2}.\nonumber
  \end{split}
\end{equation}
Substituting Eq. \eqref{3.16} into $\overline{\langle \dot x\rangle }= \int_{0}^{L}
\overline{G(x,t)}\,dx$, where $\overline{(\cdot)}$ denotes time averaging,
we obtain the average protein transport velocity,  as a function of the noise intensity $\sigma^2$ and the peptide size $N$
\begin{equation}
\label{3.17}
\overline{\langle \dot x\rangle }\approx 
\frac{L^2A^2}{2I_{10}I_{20}}\left[\frac{I_{11}}{I_{10}}-\frac{I_{21}}{I_{20}}-\frac{N L}{\sigma^2}\left(1-\frac{2I_{30}}{I_{10}I_{20}}
\right)\right].
\end{equation}

\begin{figure}[t!]
\centerline{\includegraphics[width=0.47\textwidth]{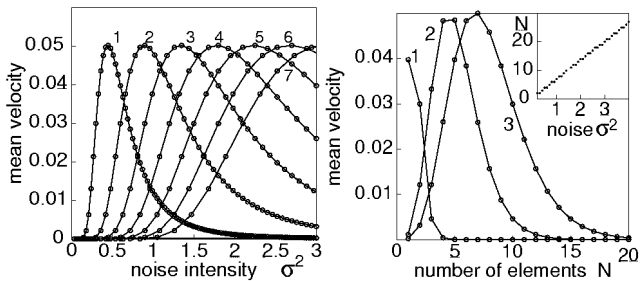}}
\caption{left: Mean velocity versus noise intensity for 
$N=1,2,..7$.
right: Mean velocity vs  $N$ for different temperatures:  $\sigma^2=0.6$(curve 1), 
$2$, and $3$ (curves 2 and 3). The inset plot shows the depth of the protein penetration 
for velocity  $0.0005$. 
}
\label{theory}
\end{figure}

The dependence of the transport velocity on the noise intensity, i.e. temperature, is shown
in Fig.~\ref{theory}~(left) for different system size $N$.
 This plot shows that protein transport 
is possible only in a certain temperature interval. Absence of noise leads to vanishing  transport
as well as too high temperature also decreases the velocity. For given temperature, the peptide size $N$ strongly influences the transport velocity, i.e., with increasing $N$ the dependence is shifted to the right and becomes more wide (curves 1-3). 
This dependence differs qualitatively for small and large temperature. 
For rather small temperature the velocity decays with increasing  $N$ (curve 1 in Fig.~\ref{theory}~(right)). 
Consequently, when a protein enters the proteasome, it will initially move with rather large velocity. Then, due to the increase of $N$ the velocity decreases, and the motion of the protein virtually terminates.
On this stage the probability of cleavage is much higher than that 
of translocation, thus, with large probability the protein will be cleaved there.
After cleavage, $N$ is significantly decreased (Fig.~\ref{work}~(right)), 
and following the velocity dependence, both fragments move rapidly to leave the proteasome.
Noteworthy, in agreement with our model, the blocking of uncleaved  proteins inside the proteasome has been observed also experimentally \cite{2001_Reid_pnas}.

Surprisingly, for large temperature the velocity depends qualitatively different on the peptide size (curves 2,3 in  Fig.~\ref{theory}~(right)), i.e., there is an optimal peptide size that corresponds to maximal velocity. In this case the protein moves accelerated while entering the proteasome until the optimal velocity is achieved. Then the velocity decreases, and again it virtually stucks.
After cleavage the behavior is quite different:
if the cleaved peptide is close to the optimal size, it leaves the proteasome rapidly.
If the cleaved fragment is too small or too large, it moves with small velocity, thus blocks the proteasome leading to low efficiency or malfunction of the proteasome.
Note that for the protein that enters the proteasome the initial small $N$ increases, and the velocity will increase, whereas for the cleaved fragment of smaller $N$ the velocity remains constant. There is experimental evidence \cite{2001_Reid_pnas} that certain proteins cannot be cleaved by the proteasome, i.e., they are inert with respect to its active sites. For such proteins we can determine the penetration depth, defined as the depth $N$ at which the propagation velocity falls below a predefined very small threshold. The inset of Fig.~\ref{theory}~(right) shows that the penetration depth increases linearly with temperature. So far, there are no experimental data to verify this prediction.

Consequently, small temperature leads to  preferably small size of the cleaved fragments, whereas larger temperature leads to domination of some specified peptide size in the proteasome output. As a hypothesis, we believe that the qualitatively different proteasome behavior with respect temperature may be responsible for temperature reaction or heat shock response
 \cite{2000_Kuckelkorn,2001_Stepanenko_mol_biology,2000_Wyttenbach_pnas}, providing immune defense in the case of some diseases. For large temperature (Fig.~\ref{theory}~(right)~curve 2,3) the protein is accelerated while entering the proteasome, hence, leading to more efficient operation of the proteasome. 

\begin{figure}[t!]
\centerline{\includegraphics[width=0.47\textwidth]{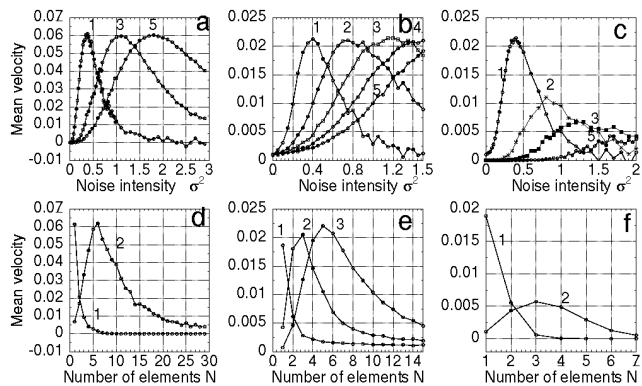}}
\caption{a,b,c:  Mean velocity versus noise intensity for different $N=$1,2,..5.
Here $A=1.15$, $\omega=0.1$.
d,e,f: Mean velocity vs. $N$ for different temperatures: d) $\sigma^2=0.3$ (1), $2$ (2); 
e) $\sigma^2=0.3$ (1), $1$ (2), and $2$ (3); f)$\sigma^2=0.5$ (1) and $1.5$ (2).
a,d: with periodic force, b,e: with colored noise, c,f: without global fluctuations.
}
\label{model1}
\end{figure}

To confirm these  results we have performed computer simulations \cite{1992_Kloeden_book} whose results agree with our analytical findings (Fig.~\ref{model1}).  
Some quantitative deviations result from taking into account only a limited number of terms
in the expansion Eq.\eqref{3.16}.

{\it Case 2:} Since collective fluctuations of the protein or peptide elements,
 represented by a periodic in time force, 
can be used only  under certain assumption, next 
 we consider the system  in the absence of deterministic periodic forces, but with
   colored noise $F(t)=\zeta(t)$, where
$\zeta(t)$ is harmonic noise with some dominating frequency (due to the molecular spectrum and oscillations of the peptide chain \cite{2000_Tama_proteins,1997_Julicher_prl}). We obtain this noise as the solution of
\begin{equation}
  \ddot \zeta +2 \delta \dot \zeta+\omega_0^2 \zeta = \xi^\prime,  
\end{equation}
where $\xi^\prime$ is Gaussian noise $\langle \xi^\prime(t)\xi^\prime(t^{\prime})\rangle=\sigma^2_a\delta(t-t^{\prime})$,
and parameters $\delta=0.01$, $\omega_0=0.1$, $\sigma^2_a=10^{-4}$ regulating the width, the dominating frequency of
 the power spectrum, and the intensity of $\zeta(t)$.
The results  show qualitatively the same behavior as in the case
of the periodic force (Fig.~\ref{model1} (b,e)). Hence,  if exclusively noisy fluctuations are present
in the system, the generic behavior described above, persists and leads to the same temperature and
peptide size dependences.

{\it Case 3:} Next we consider the case of no collective fluctuations, $F(t)=0$, i.e., there 
is no eigen collective mode in the peptide oscillations.
The local noise of each interaction center 
is presented now by the sum of uncorrelated colored and white noise components: $f_i(t)=\zeta_i(t)+\xi_i(t)$.  The presence of the colored noise is necessary to generate nonequilibrium
 fluctuations, that
is a necessary condition for directed transport. In this case the motion of the protein or peptide
is described by
\begin{eqnarray}
\dot x = - \frac{\partial U(x)}{\partial x}+\frac{1}{N}(\zeta_1+...+\zeta_N+\xi_1+...\xi_N),
\label{eq:1}
\end{eqnarray}
with $\zeta_i(t)$ and $\xi_i(t)$ as above. In general, even in this case the behavior is similar, but
increase of the peptide size leads not only to a shift of the velocity dependence towards large noise intensity, but also to a decrease of its maximal values (Fig.~\ref{model1}~(c)). This is due to the fact that nonequilibrium fluctuations originating from different interaction centers act
occasionally in converse direction. For small temperature (Fig.~\ref{model1}~(f), curve 1) the transport decreases with increasing peptide size $N$, whereas for large temperature the rate depends nonmonotonously on $N$ (curve 2). The optimal velocity in the latter case is smaller, as for the case of collective  fluctuations.

In summary, we investigated a model for active protein transport inside the proteasome analytically and numerically. The size of the peptide as well as temperature influence the transport rate significantly in agreement with experimental results. Following our predictions, protein transport occurs only in the certain interval of temperature. The size dependence of the transport velocity leads to a preferred fragment (peptid) size since for such sizes the probability of cleavage exceeds the probability of further translocation significantly. Under certain conditions, uncleaved proteins may get stuck inside the proteasome, thus leading to decrease of its efficiency which is in good agreement with experimental observations \cite{2001_Reid_pnas}. 
Calculating the average velocity, we have considered only the periodic constituent of the protein-proteasome potential. Slight inhomogeneity (nonperiodicity) of this potential will lead to fluctuations of the velocity around its average leaving the qualitative behavior of dependencies on the peptide size intact.

Varying temperature, the model predicts qualitatively different transport regimes, thus, possibly explaining the mechanism of 
temperature reaction or {\em heat shock response} as it is observed for certain diseases. For larger temperature, the protein is accelerated while entering the proteasome, moreover, there is a preferred fragment size, that should dominate the proteasome output. We emphasize that our model is based only on two assumptions: spatial asymmetry of the protein chain and presence of nonequilibrium fluctuations, which seem to be certainly fulfilled in real proteasome machines. 

We thank C. Fr\"ommel, H. Holzh\"utter, and R. Prei{\ss}ner for helpful discussion.


\begin{thebibliography}{10}
\bibitem{asymmetry}
While $N$ wears a positive partial charge (via the attached $H$) and $C$ is negative (via $O$), the third component of the backbone, $CH$ with the attached residue is essentially neutral. Its charge is determined by the character of the residue which is attributed to the irregular part of the backbone. The sequence $\dots (-)\to(+)\to(0)\to(-)\to(+)\to(0)\dots$ gives rise to an asymmetric interaction potential as sketched in Fig. \ref{potential}.
\end{thebibliography}
\end{document}